\documentclass[a4paper]{article}

\usepackage[english]{babel}
\usepackage[utf8x]{inputenc}
\usepackage[T1]{fontenc}

\usepackage[a4paper,top=3cm,bottom=2cm,left=3cm,right=3cm,marginparwidth=1.75cm]{geometry}

\usepackage{amsmath}
\usepackage{graphicx}
\usepackage{authblk}
\usepackage[colorinlistoftodos]{todonotes}
\usepackage[colorlinks=true, allcolors=blue]{hyperref}

\title{Smith-Purcell radiation driven by plasma accelerated beam}
\author{P. Zhang}
\affil{School of Electronic Science and Engineering, University of Electronic Science and Technology of China}
\begin{document}
\maketitle

\begin{abstract}
The Smith Purcell radiation as one promising THz radiation source has been studied for decades. Various structure designs and beam configurations have been proposed. In this paper, we report our recent exploration of Smith Purcell radiation driven by electron beam accelerated by plasma wakefield.
\end{abstract}

\section{Introduction}
Smith-Purcell radiation (SPR) is a radiation emitted from the collective oscillations of induced charge by charged particles passing over a periodic surface. It was observed for the first time in 1953 \cite{smith1953}. Since then many research groups reported SPR observation from visible light to millimeter wavelength\cite{Kube2002,Brownell1997,Doucas1992,Woods1995}. However, these SPR observed were incoherent. Starting from 1990s, SPR were discussed both as a THz radiation source and an electron beam longitudinal profile diagnostic\cite{Nguyen1997,Lampel1997}.

\begin{figure}
\centering
\label{fig:grating}
\includegraphics[width=0.7\textwidth]{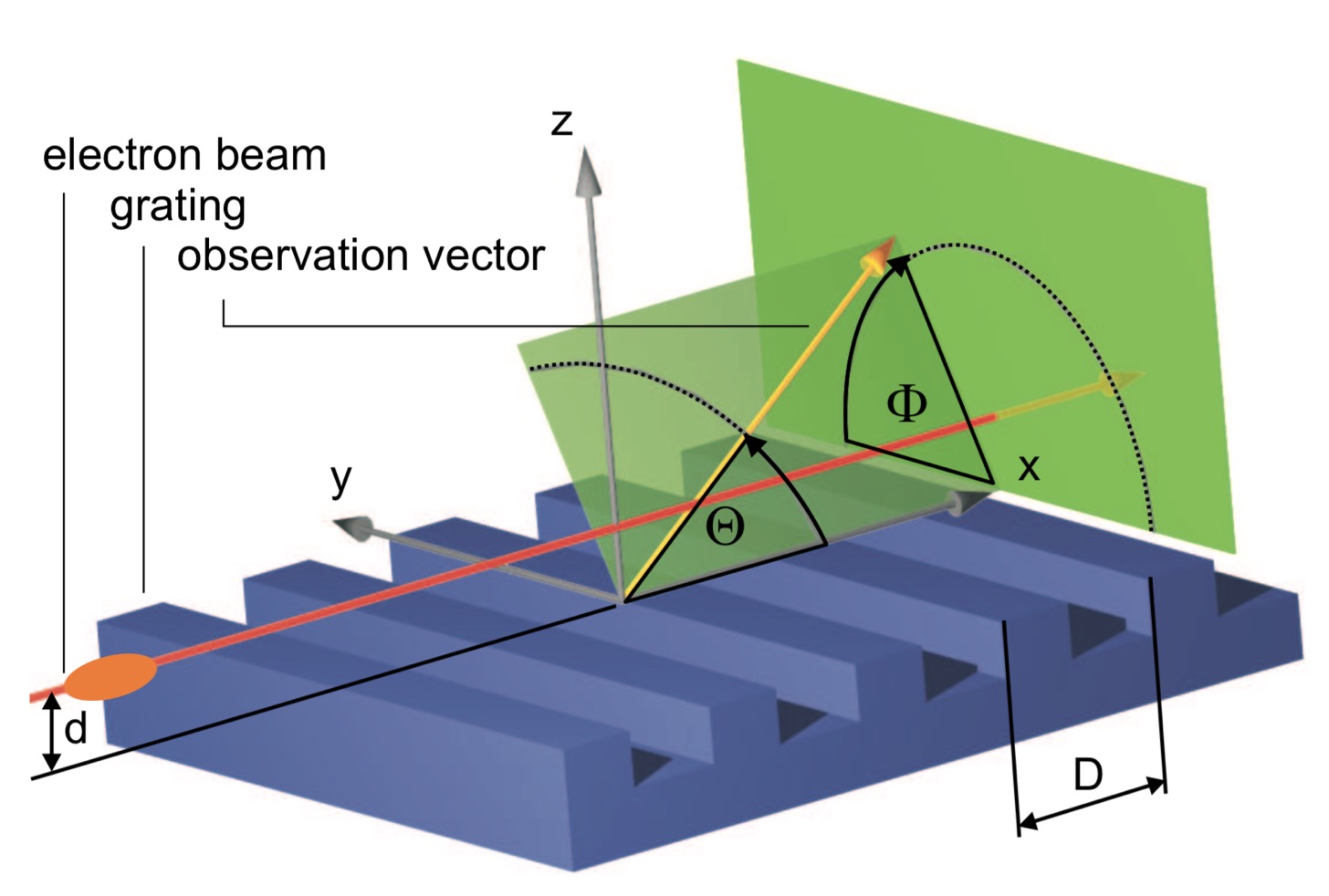}
\caption{Geometry of typical rectangular grating structure as shown in \cite{Backe2006}. Electron bunch (red) travels in $x$ direction above the surface of structure at the distance of $d$. The groves are in $y$ direction with period of $D$. The polar angle with respect to $x$-axis is $\Theta$ and the azimuthal angle with respect to $y$-axis is $\Phi$.}
\end{figure}

There are many theories to explain the radiation in metallic gratings\cite{Berg1973,Brownell1998,Potylitsyn1998}. One approach was discussed in di Francia's work. The radiation emission mechanism can be seen as the diffraction of electron field by grating structure. The angular distribution of the number of radiation photons in unit solid angle is expressed as \ref{eq:angdis}

\begin{equation}
\label{eq:angdis}
\frac{dN}{d\Omega} = \alpha|n|N_w\frac{sin^2\Theta sin^2\Phi}{(1/\beta - cos\Theta)^2}|R_n|^2e^{-\frac{d}{h_{int}}\xi(\Theta, \Phi)}
\end{equation}

where $\xi(\Theta, \Phi) = \sqrt{1+(\beta\gamma sin\Theta cos\Phi)^2}$, $\alpha$ is the fine-structure constant, $N_W$ is the number of grating periods, $\beta=v/c$ is electron's normalized velocity, $\Theta$ and $\Phi$ are emission angles. The interaction length $h_{int}$ is $\beta\gamma\lambda/4\pi$ where $\gamma=(1-\beta^2)^{-1/2}$ is the Lorentz factor. 

As shown in Equation \ref{eq:angdis}, the increase of distance from beam to surface of grating structure $d$ causes the decrease of intensity exponentially. The minimum distance limited by the normalized transverse emittance of beam $\epsilon_{N}$ is $d_{min}=\sqrt{L\frac{\epsilon_N/\pi}{\beta\gamma}}$ derived in \cite{Backe2006}. Therefore, low emittance beam can generate much higher radiation at the same distance due to the exponential scaling.

On the other hand, the wavelength of radiation is \ref{eq:wv} where $D$ is the grating period. This indicates that higher energy electron beam should generate shorter wavelength radiation. In Smith and Purcell's experiment, 250-300 KeV electron beam was employed and radiation is in visible light wavelength. In 1998, Urata et al observed first superradiant Smith-Purcell radiation at wavelength of 491 $\mu m$ with a 35 KeV electron beam\cite{Urata1998}. 1.44 MeV beam was used in experiment discussed in \cite{Backe2006} to produce Smith-Purcell radiation in the wavelength region between 30-300 $\mu m$ to fill the THz gap.

\begin{equation}
\label{eq:wv}
\lambda=\frac{D}{|n|}(1/\beta-cos\Theta)
\end{equation}

The intersection of two beam configuration dimension requires high energy  and ultra-low emittance beam. Advanced accelerator technologies have the potential to fulfill this requirement. One path is plasma wakefield acceleration. Since Tajima and Dawson \cite{dawson1979} proposed the concept of injecting laser through plasma and creating perturbation, it has been explored and demonstrated with many variations. The accelerating gradient of plasma wakefield accelerator can be orders of magnitude higher than conventional radio-frequency (RF) accelerators. As plasma-based acceleration has made tremendous progress in last decades, higher energy electron beam nowadays can be achieved in table-top-scale distance. Depends on the wake is driven by electron bunch or laser pulse, the scheme is called plasma wakefield accelerator (PWFA) \cite{joshi2006} and laser wakefield accelerator (LWFA) \cite{Esarey2009} respectively. Especially, when driver bunch density $n_b$ is much higher than plasma density $n_p$, plasma electrons in the path of driver will be expelled to side leaving an ion column behind the driver\cite{Rosenzweig1991} which is so called "blow-out regime". In this scenario, longitudinal electric field strength $E_z$ is proportional to $n_p^{1/2}$ which indicates higher density plasma yields higher accelerating gradient.

Besides improving accelerating gradient, accelerator scientists also studied reducing beam emittance and increasing beam brightness by varying injection scheme. One promising scheme nicknamed "Trojanhorse" was proposed by Hidding et al\cite{Hidding2012,Hidding2013,Hidding2014}. The idea is to use a driver bunch (electron bunch or laser pulse) to create a blowout regime in gas mixture(e.g. $\mathrm{H_2}$ and He), a timing-synchronized focused laser pulse is injected to ionize high-ionizing threshold gas such as He and release electrons. These electrons will be trapped inside the ion bubble and form the witness bunch while experiencing acceleration. Due to the trivial initial momentum of tunnel ionization and focusing transversal wakefield, the scheme has the potential to deliver ultra-low emittance ($\sim$0.01 mm mrad) beam\cite{Xi2013}. The proof-of-concept experiment has been carried out at FACET in SLAC\cite{Wittig2015,Wittig2016,Manahan2016,Knetsch2014}. It is worth to mention that in this experiment driver bunch and ionizing laser pulse have to be synchronized to sub-hundred femtosecond level\cite{Xi2016}. This is achieved by employing Electro-Optic Sampling (EOS) technique. When an electron bunch passes by crystal such as ZnTe or GaP, the strong electric field modifies the refractive index of crystal thus rotating the polarization of probe laser which can be captured by CCD camera. While "Trojanhorse" beam is a strong candidate to serve the next generation X-FEL, it could be interesting to drive Smith-Purcell radiation alternatively. 

\section{Simulation}

\begin{figure}
\centering
\label{fig:40keV}
\includegraphics[width=0.7\textwidth]{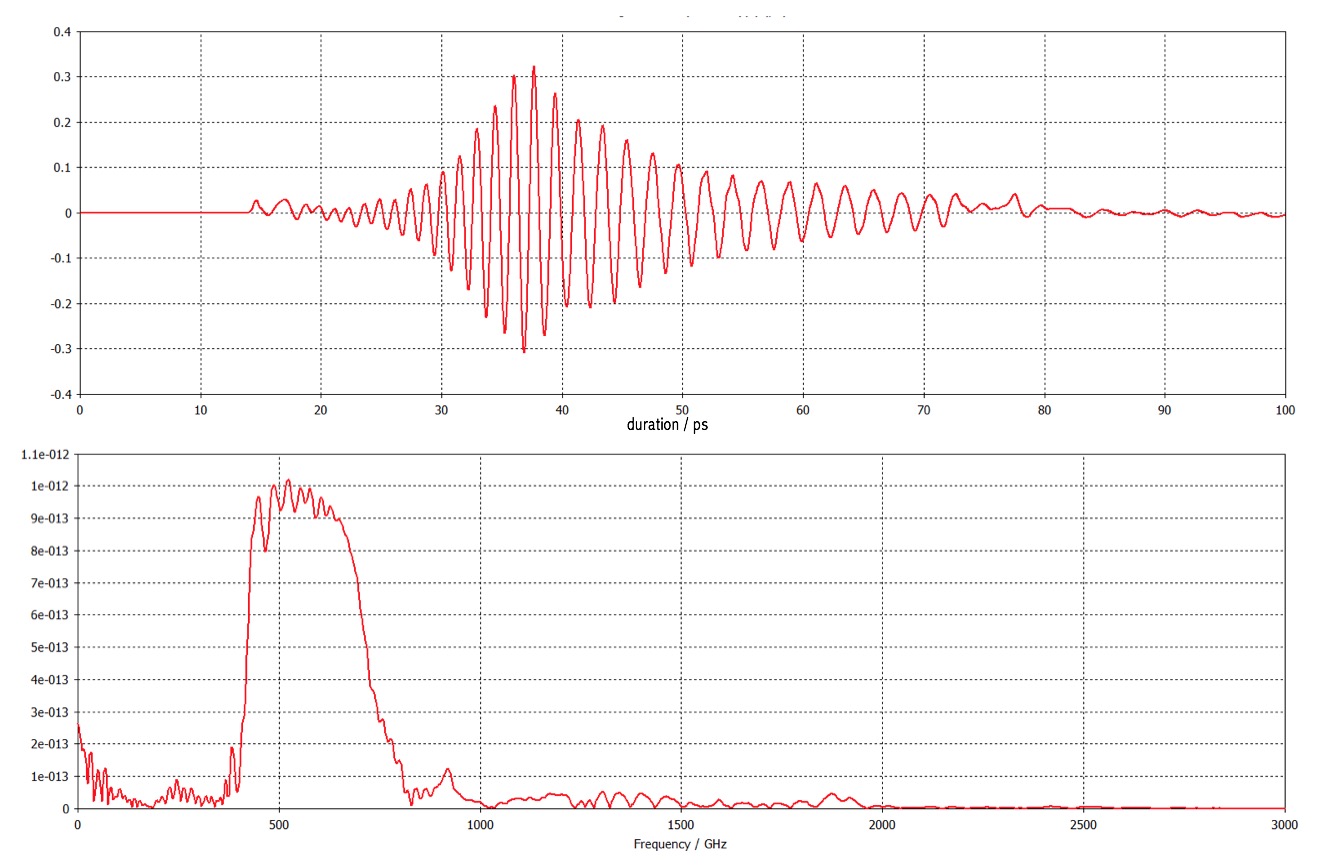}
\caption{Radiation wave in time domain (top) and in frequency domain (bottom) for 40 KeV driver beam.}
\end{figure}

\begin{figure}
\centering
\label{fig:1GeV}
\includegraphics[width=0.7\textwidth]{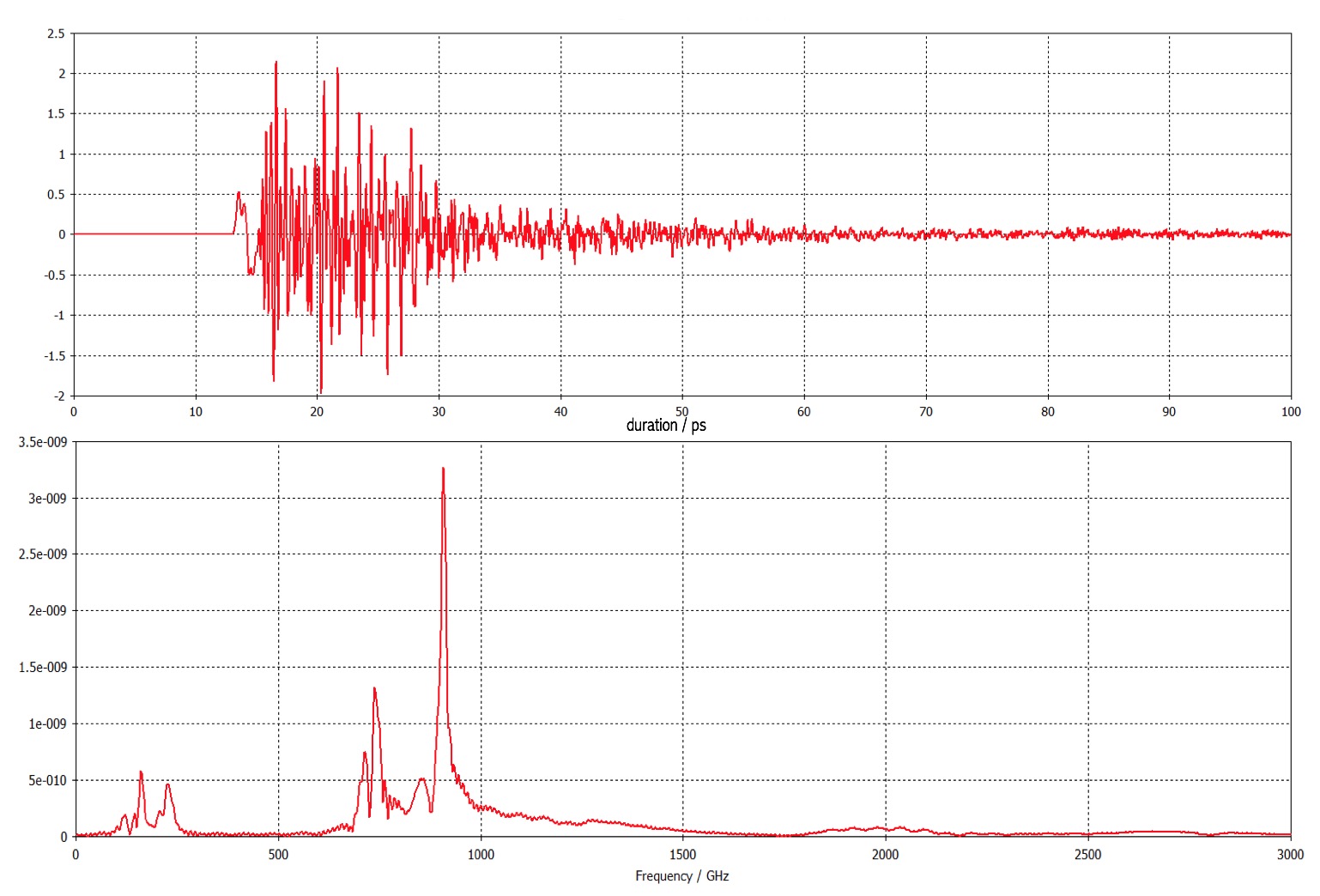}
\caption{Radiation wave in time domain (top) and in frequency domain (bottom) for 1 GeV driver beam.}
\end{figure}

We carried out simulations to demonstrate our theories. The first step is to compare the effect of driver beam energy on Smith-Purcell radiation frequency. In simulation, we employed rectangular metallic grating with period 0.2 mm. The depth of grating groove is 0.1 mm. We compared the radiation both in time domain and in frequency domain for 40 KeV and 1 GeV beam. The simulation result is shown in Figure \ref{fig:40keV} and Figure \ref{fig:1GeV}. When electron beam energy is 40 KeV, the wave form in time domain is more regular while 1 GeV beam is more fluctuating which indicates there are high frequency components interplaying. This is confirmed in frequency spectrum. The radiation peak of 40 KeV beam has about 400 THz width and is centered around 600 THz. However, the width of 1 GeV beam case is much narrowed which implies the radiation is quasi-mono-chromatic. This advantage can be used to produce extremely narrow bandwidth radiation. Moreover, the peak is shifted to around 900 THz which proves higher energy driver beam emits higher frequency/shorter wavelength radiation.

\begin{figure}
\centering
\label{hundredkeV}
\includegraphics[width=0.7\textwidth]{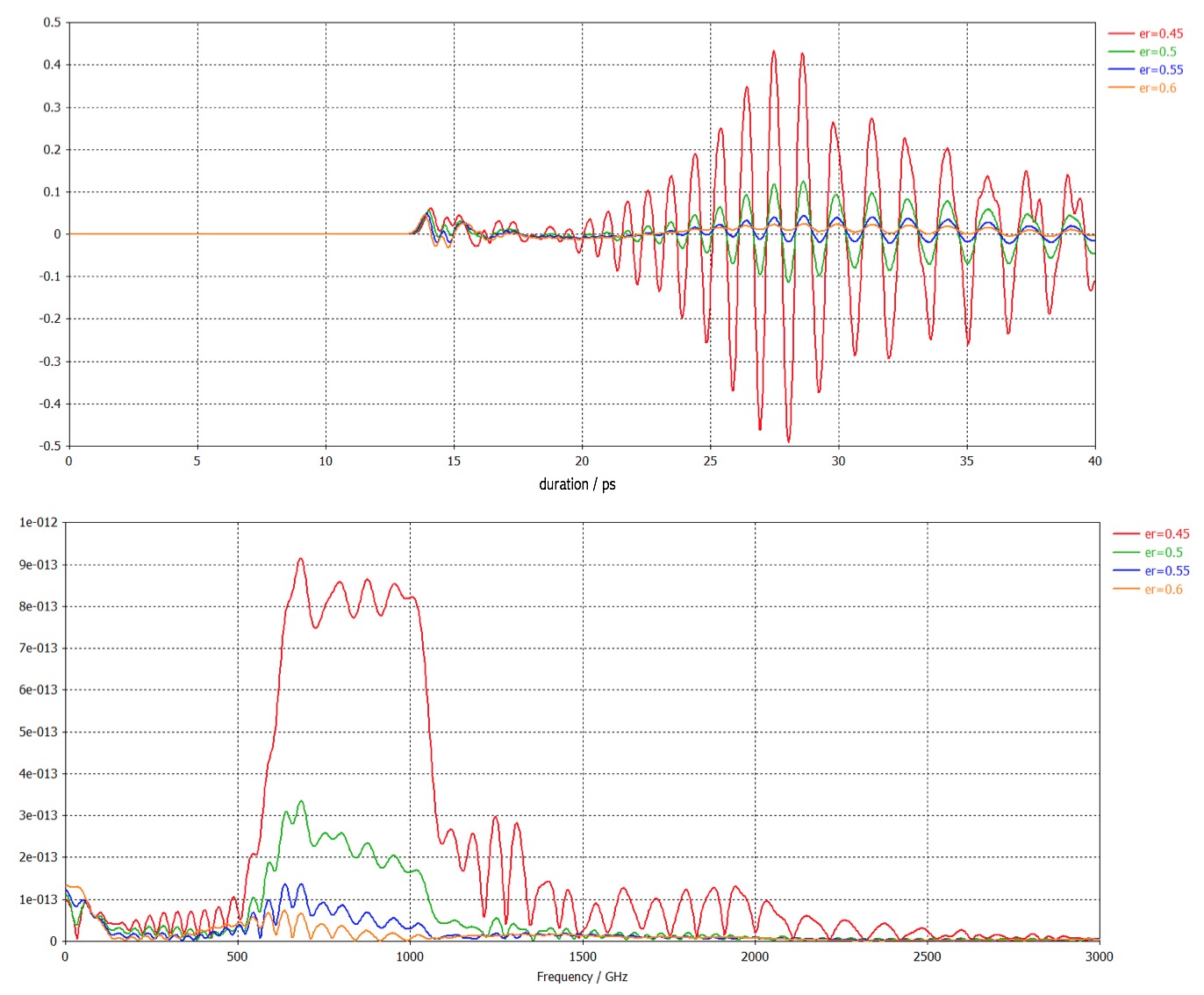}
\caption{Radiation wave in time domain (top) and in frequency domain (bottom) for 1 00 KeV driver beam. The distance between electron beam and surface of structure is 0.45 mm(red), 0.5 mm(green), 0.55 mm(blue), 0.6 mm(brown) respectively.}
\end{figure}

Regarding the relationship between beam-surface distance and radiation strength, since electron beam from plasma wakefield acceleration has ultra-low emittance and allows us to inject beam closer to the surface of grating structure, we also did simulation to study the radiation amplitude with different distance. The same grating structure was used in simulation. The energy of electron beam is set to 100 KeV. The distance between electron beam and surface of structure is from 0.45 mm to 0.6 mm at the step of 0.05 mm. The wave form in time domain and frequency spectrum are shown in Figure. As the distance increases, the strength of radiation decreases drastically. This provides practical guidance for experiments.  

In conclusion, we discussed the influence of driver beam parameters(beam energy and beam emittance) in the Smith-Purcell radiation quality and thus proposed to employ accelerated beam from plasma accelerator. From theories developed earlier, the radiation wavelength is determined by driver beam energy and the simulation confirmed that higher driver beam energy can emit shorter wavelength radiation. The low beam emittance gives us more tolerance of distance between driver beam and grating surface structure from experiment perspective. Closer distance can increase radiation amplitude significantly. In the future, as more sophisticated grating structures \cite{Zhang2015,Zhang2017} are proposed the efficiency of radiation will be further improve.

\bibliographystyle{unsrt}
\bibliography{sample}

\end{document}